\documentstyle[12pt]{article}
\if@twoside  m
\else
\fi
 1
\font\gt = eufm10 scaled \magstep 1
\font\sr = msbm10 scaled \magstep 1

\def\Tl{${\cal T}^\lambda\ $}

\def\Hl{${\cal H}_\lambda\ $}

\def\Uh{{\cal U}_h}

\def\id{{\rm id}}

\def\vaca(\langle\cdot\rangle\}
\def\diag{{\rm diag}\,}

\def\gtg{{\gt g}}
\def\gtk{{\gt k}}
\def\gtb{{\gt b}}
\def\gtn{{\gt n}}
\def\gth{{\gt h}}
\def\gtp{{\gt p}}

\def\sgn{{\rm sgn}}

\begin{document}
\begin{titlepage}
\begin{flushright}
CERN-TH.7489/94\\
LBL-36321\\
hep-th/9411038\\
\end{flushright}
%\begin{flushleft}
%October 1994
%\end{flushleft}
\vspace{.1cm}
\begin{center}
{\Large\bf On Fock Space Representations of\\
Quantized Enveloping Algebras Related to\\
 Non-commutative Differential Geometry.} \\
%{\Large \bf}\\
\vspace{1.3cm}
{ B. Jur\v co}\\
{\em CERN, Theory Division}\\
{\em CH-1211 Geneva 23, Switzerland}\\
\vspace{.5cm}
{M. Schlieker\footnote{This work was supported in part by the
Director, Office
of Energy
Research, Office of High Energy and Nuclear Physics, Division of High
Energy Physics of the U.S. Department of Energy under Contract
DE-AC03-76SF00098 and in part by the National Science Foundation
under grant PHY-90-21139.}$^,$\footnote{Supported in part by a
Feodor-Lynen
Fellowship.}}\\
{\em Theoretical Physics Group}\\
{\em Lawrence Berkeley Laboratory}\\
{\em University of California}\\
{\em Berkeley, CA 94720, USA}

\end{center}
\vspace{1cm}
\begin{center}
{\bf Abstract}
\end{center}
{\small In this paper we construct explicitly natural (from the
geometrical
point of view) Fock space
representations (contragradient Verma modules) of the quantized
enveloping
algebras. In order to do so, we start from the Gauss decomposition
of the quantum group and introduce the differential operators on the
corresponding $q$-deformed flag manifold (asuumed as a left comodule
for the
quantum group) by a projection to it of the right action of the
quantized
enveloping algebra on the quantum group.
Finally, we express the representatives of the elements of the
quantized
enveloping algebra corresponding to the left-invariant vector fields
on the
quantum group as first-order differential operators on the
$q$-deformed flag
manifold.}
\vspace{2cm}\\
CERN-TH.7489/94\\
October 1994
\end{titlepage}

\newpage

\setcounter{footnote}{0}
\setcounter{page}{1}
{\thispagestyle{empty}

\section{Introduction}
Let $G$ denote a simple and simply connected complex Lie group and $K
\subset
G$ its compact form.
The purpose of this paper is to construct an explicit representation
of the quantized enveloping algebra $\Uh$(\gtk) of the quantum group
$K_q$ in
terms of (local) holomorphic coordinates and differential operators
on the
homogeneous space $(K_0\backslash K)_q=(P_0\backslash G)_q$. This
extends
unambiguously to a representation of $\Uh$(\gtg). Starting point for
this
construction is the Gauss decomposition of the canonical element of
the
quantum double of $G_q$ (which yields the Gauss decomposition of the
vector
corepresentation) as described in \cite{J-St}. Using the projection
from
$G_q$ to $(P_0\backslash G)_q$ we introduce partial differential
operators on
the homogeneous space. The representation of the elements of the
universal enveloping algebra in terms of these differential operators
and holomorphic functions can be viewed as a natural Fock space
representation of $\Uh$(\gtg) ($\Uh$(\gtk)) (the contragradient of
the Verma
module). The explicit fomulas are presented in the case $G=SL(N)$ and
$P_0=B_-$
with $B_-$ being the Borel subgroup of lower triangular matrices.
However, we
hope that from our exposition it is clear that the general case can
be treated
similarly. We assume a generic value of $q=e^{-h}$, which becomes
real, while
referring to the compact forms.

Let us mention that there are already many papers devoted to the
subject. We
mention just a few \cite{Awata et al}, \cite{sr}, which seem to be
most closely
related to our approach.
Nevertheless, we wish to stress the geometric origin of our
construction, which
employs the Borel--Weil-like description of the irreducible
representations of
the quantized enveloping algebra $\Uh$(\gtk). It follows from
$\cite{J-St}$
that the representatives of the elements of $\Uh$(\gtk) ($\Uh$(\gtg))
corresponding to the left-invariant vector fields can be described
completely
in terms of non-commutative differential geometry on the quantum
group $K_q$
($G_q$). What we are doing here is essentially an
 explicit restriction of the non-commutative differential calculus on
$K_q$
($G_q$) to the algebra of holomorphic functions on the homogeneous
space
$(K_0\backslash K)_q=(P_0\backslash G)_q$ and finally expressing all
in terms
of this only. This is non-trivial mainly because the projection
(\ref{proj}) is
no longer an algebra homomorphism as in the classical
case.\vspace{.5cm}\\

\section{Preliminaries, notation}
In this section we repeat some of the results described in
\cite{J-St}. For the
general construction of the quantum double and its relation to
quantum groups, we refer to \cite{Dr}, \cite{F-R-T}, \cite{W}.

Let $\Uh$(\gtg) be the quantized enveloping algebra related to a
simple Lie algebra {\gtg} and $F_q(G)$ the dual Hopf algebra of
quantized functions on the corresponding simple Lie group $G$.

%\cite{Dr}, \cite{Ji}, \cite{F-R-T}, \cite{W}.
Let further $\rho$ be the canonical element
$$
\rho=\sum x_s\otimes a_s\in \Uh(\mbox{\gtg})^{op\Delta}\otimes
F_q(G)\, ,
$$
with $\{ x_s\}$ and $\{ a_s\}$ being mutually dual bases. Its basic
properties are ($S$ is the antipode, $\Delta$ the
comultiplication)
%\cite
$$
\rho^{-1}=(\id\otimes S)\rho\, ,
$$
\begin{equation}
(\Delta\otimes\id)\rho=\rho_{23}\rho_{13},\
(\id\otimes\Delta)\rho=\rho_{12}\rho_{13}\, .\label{bp}
\end{equation}

Denote by {\gt b}$_\pm\subset\,${\gt g} the Borel subalgebras and by
{\gt h}$=\,${\gt b}$_+\cap\,${\gt b}$_-$ the Cartan subalgebra.

Fixing a maximal Weyl element, one orders the set $\Delta^+$
of positive roots as $(\beta_1,\dots,\beta_d),\ d=|\Delta^+|$. To
each
root $\beta_j$ there are related elements $E(j)\in\Uh$(\gtb$_+)$ and
$F(j)\in\Uh$(\gtb$_-)$, so that the elements
\begin{equation}
E(d)^{n_d}\dots E(1)^{n_1}H_l^{\ m_l}\dots H_1^{\ m_1}\, \label{cb}
\end{equation}
($n_i,\ m_i\in\,${\sr Z}$_+$), form a basis in $\Uh$(\gtb$_+)$
\cite{kirillov-Reshetikhin}, \cite{Levendorskii-Soibelman}. The
vectors
$H_i$ can be replaced by any elements forming a basis in \gth.\,
A similar
assertion is valid also for $\Uh$(\gtb$_-)$. In the limit
$h\downarrow 0$
the elements $E(j)$ and $F(j)$ become the root vectors
$X_{\beta_j}\in\,$\gtn$_+$ and $X_{-\beta_j}\in\,$\gtn$_-$,
respectively. We recall that the universal $R$-matrix can be written
in the form
\cite{kirillov-Reshetikhin}, \cite{Levendorskii-Soibelman}
\begin{equation}
R^u=\exp_{q_d}\bigl(\mu_d\, F(d)\otimes E(d)\bigr)\dots
\exp_{q_1}\bigl(\mu_1\, F(1)\otimes E(1)\bigr)\, \exp(\kappa)\, ,
\end{equation}
where $\exp_q$ are the $q$-deformed exponential functions, $\mu_j$
are
some coefficients depending on the parameter $h$, and $\kappa$ is
some
element from $\Uh$(\gth)$\otimes\Uh$(\gth).

We make use
of the fact that
$\Uh$(\gtg)$^{op\Delta}$ is a factor algebra of
$\Uh$(\gtb$_-)^{op\Delta}\otimes_{\rm twist}\Uh$(\gtb$_+)^{op\Delta}$
and
$F_q(G)\simeq\Uh$(\gtg)$^\ast$ is a subalgebra in
$\Uh$(\gtb$_+)^{op\cdot}\otimes\Uh$(\gtb$_-)^{op\Delta}$. The
canonical
element $\tilde\rho$ in
\begin{equation}
\Bigl(\Uh(\hbox{\gtb}_-)^{op\Delta}\otimes_{\rm twist}
\Uh(\hbox{\gtb}_+)^{op\Delta}\Bigr)\otimes\Bigl(
\Uh(\hbox{\gtb}_+)^{op\cdot}\otimes\Uh(\hbox{\gtb}_-)^{op\Delta}\Bigr)
\end{equation}
can be decomposed as follows
\cite{FRT}
\begin{eqnarray}
\tilde\rho & = & \sum (e_j\otimes e^k)\otimes(f^j\otimes f_k) &
\cr\cr
& = & \sum (e_j\otimes 1\otimes f^j\otimes 1)\cdot
(1\otimes e^k\otimes 1\otimes f_k) \cr\cr
& = & \tilde R_{13}\tilde R'_{24}\, .
%& \cr\cr
 \label{ce}
\end{eqnarray}
Here $\{e_j\},\ \{e^k\},\ \{f^j\}$ and $\{f_k\}$ stand for bases in
the
corresponding factors, $\{e_j\}$ and $\{f^j\}$ are dual and the same
is
assumed of $\{e^k\}$ and $\{f_k\}$; the dot in the third member of
equalities (\ref{ce}) indicates multiplication in the double and
$\tilde R'$ is
obtained from $\tilde R$ by reversing the order of multiplication. To
express
$\rho$ we shall again use bases of the type (\ref{cb}). In the
notation adopted here,
the elements
$F(j),\ E(j),\ \tilde E(j)$ and $\tilde F(j)$ belong in this order to
the
individual factors in (\ref{ce}). Factorizing off the redundant
Cartan
elements we have
\cite{J-St}
\proclaim Proposition 1.
The canonical element for the quantum double
$\Uh(g)^{or\Delta}\otimes F_q(G)$
has the form
\begin{eqnarray}
\rho & = & \exp_{q_d}\bigl(\mu_d\, F(d)\otimes\tilde E(d)\bigr)\dots
\exp_{q_1}\bigl(\mu_1\, F(1)\otimes\tilde E(1)\bigr)\, \exp(\kappa)
&  \cr\cr
& & \times\exp_{q_1}\bigl(\mu_1\, E(1)\otimes\tilde F(1)\bigr)\dots
\exp_{q_d}\bigl(\mu_d\, E(d)\otimes\tilde F(d)\bigr)\, .
% &
% \cr\cr
\end{eqnarray}

Let further  $\Pi_0$ denote any subset of the set of simple roots
$\Pi$ and let us denote by $\Uh$(\gtg$_0$) the Hopf subalgebra in
$\Uh$(\gtg) generated by all Cartan elements $H_i$, and only by those
elements $X^\pm_i$ for which $\alpha_i\in\Pi_0$. Similarly we shall
denote by $\Uh$(\gtp$_0$) the Hopf subalgebra in $\Uh$(\gtg)
generated
by all $H_i,\ X^-_i$ and those $X^+_i$ for which $\alpha_i\in\Pi_0$.
The maximal Weyl element can
be chosen such that there exists $p\in\,${\sr Z}$_+$, $p\leq d$, such
that
the vectors $X_{-\beta_1},\dots, X_{-\beta_d}$, $H_1,\dots, H_l$,
$X_{\beta_1},\dots,X_{\beta_p}$
form a basis of \gtp$_0$. Then $X_{\beta_{p+1}},\dots,X_{\beta_d}$
form a basis of a nilpotent subalgebra \gtn$_0$ and
\gtg$=$\gtp$_0\oplus$\gtn$_0$.  This means that all elements $F(j)$
belong to
$\Uh$(\gtp$_0)$, while $E(j)$ belongs to $\Uh$(\gtp$_0)$ only for
$j=1,\dots,p$.
Notice that in the generic case $\Pi_0=\emptyset$ and hence $p=0,$
\gtp$_0=\,$\gtb$_-$
and \gtn$_0=\,$\gtn$_+$.

As it is easy to see, $\Uh$(\gtg$_0$) is again a quasitriangular
Hopf algebra with the universal $R$-matrix $Q^u$ given by
\begin{equation}
Q^u=\exp_{q_p}\bigl(\mu_p\, F(p)\otimes E(p)\bigr)\dots
\exp_{q_1}\bigl(\mu_1\, F(1)\otimes E(1)\bigr)\, \exp(\kappa)\, .
\end{equation}

The canonical element $\rho$ can be written as a product
\begin{eqnarray}
\rho & = & \Lambda {\cal Z}\, ,\quad \mbox{where} &  \cr\cr
\Lambda & = &\,
\exp_{q_d}\bigl(\mu_d\, F(d)\otimes\tilde E(d)\bigr)\dots
\exp_{q_1}\bigl(\mu_1\, F(1)\otimes\tilde E(1)\bigr)\, \exp(\kappa) &
\cr\cr
& & \times\exp_{q_1}\bigl(\mu_1\, E(1)\otimes\tilde F(1)\bigr)\dots
\exp_{q_p}\bigl(\mu_p\, E(p)\otimes\tilde F(p)\bigr)\, , & \cr\cr
{\cal Z} & = & \,
\exp_{q_{p+1}}\bigl(\mu_{p+1}\, E(p+1)\otimes\tilde F(p+1)\bigr)\dots
\exp_{q_d}\bigl(\mu_d\, E(d)\otimes\tilde F(d)\bigr)\, .\label{Z}
% & \cr\cr
\end{eqnarray}
Let us also denote
\begin{equation}
A= \exp(\kappa)\exp_{q_1}\bigl(\mu_1\, E(1)\otimes\tilde
F(1)\bigr)\dots
\exp_{q_p}\bigl(\mu_p\, E(p)\otimes\tilde F(p)\bigr)\, ,
\end{equation} so that $A{\cal Z}$ can be identified with the
canonical element
of the quantum double $\Bigl(
\Uh(\hbox{\gtb}_+)^{op\Delta}\otimes\Uh(\hbox{\gtb}_-)^{op\Delta}\Bigr
)
$; $A$ itself is the canonical element of the quantum double of
$\Bigl(
\Uh(\hbox{\gtb}_{0+})^{op\Delta}\otimes\Uh(\hbox{\gtb}_{0-})^{op\Delta
}
\Bigr)$,
where $\Uh(\hbox{\gtb}_{0-})$ is the Hopf subalgebra in
$\Uh(\hbox{\gtb}_{-})$ generated by all $H_i$ and those $X^-_i$ for
which $\alpha_i\in\Pi_0$. The projections
$P_{\pm}:\,\Uh(\hbox{\gtb}_{\pm})\rightarrow \Uh(\hbox{\gtb}_{0\pm})$
which send $E(j)$ and
$F(j)$ for $j=k+1,...,d$ to $0$ are Hopf algebra homomorphisms
\cite{J-St}.
It also holds that $(id \otimes P_-)R^u=(P_+ \otimes id)R^u = Q^u$.
As a
simple consequence of the above-mentioned facts we have
\proclaim Proposition 2.
It holds that
\begin{eqnarray}
A_{23}^{-1}{\cal Z}_{13}A_{23}=(Q^u_{12})^{-1}{\cal Z}_{13}Q^u_{12}
\, ,
\end{eqnarray}
\begin{eqnarray}
(\Delta\otimes id){\cal Z}= A_{23}^{-1}{\cal Z}_{13}A_{23}{\cal
Z}_{23}
=(Q^u_{12})^{-1}{\cal Z}_{13}Q^u_{12}{\cal Z}_{23} \, ,
\end{eqnarray}
\begin{eqnarray}
R^u_{12}(Q^u_{12})^{-1}{\cal Z}_{13}Q^u_{12}{\cal
Z}_{23}=(Q^u_{21})^{-1}{\cal Z}_{23}Q^u_{21}{\cal Z}_{13}R^u_{12} \,
. \label{cr}
\end{eqnarray}

Here $\Delta$ means the original comultiplication in $\Uh$(\gtg),
contrary to  (\ref{bp}).\vspace{.5cm}\\

Let $\tau$ designate the irreducible representation of $\Uh$(\gtg)
corresponding to the vector corepresentation $T$ of $F_q(G),\
T=(\tau\otimes\id)\rho$. As in
\cite{J-St}
we use the entries of the matrix $Z=(\tau\otimes id){\cal Z}$ as
(local) non-commutative coordinates on the $q$-deformed
homogeneous space (generalized flag manifold) $(P_0\backslash G)_q$.
We shall denote the algebra generated by these
by ${\cal C}$. Applying $\tau \otimes \tau$ to eq. (\ref{cr})
we get
the commutation relations
\cite{J-St}}
\begin{equation}
R_{12}Q_{12}^{-1} Z_{13}Q_{12} Z_{23}=Q_{21}^{-1}Z_{23}Q_{21}
Z_{13}R_{12}\,,\label{comm}
\end{equation}
where $R$ and $Q$ are used to denote the universal $R$-matrices $R^u$
and $Q^u$
in the vector representation $\tau$. Let us note that (\ref{comm})
are formally
of the same form as the defining relations of a quantum braided group
\cite{Ma}, \cite{Hl}. \vskip 0.5cm
Using the universal element ${\cal Z}$ we now introduce a mapping
(which is an algebra homomorphism in the classical case) from the
algebra of quantized functions on $G$ to the  functions on the
$q$-deformed
homogeneous
space ${\cal C}$:
\begin{equation}
\Gamma : F_q(G) \rightarrow {\cal C}:
a \mapsto (\langle a,\cdot\rangle\otimes id){\cal Z}\,. \label{proj}
\end{equation}
Unlike the classical case, the mapping $\Gamma$ is not an algebra
homomorphism on $F_q(G)$ (this is obvious from (\ref{comm})) but the
following properties are sufficient for our construction.
The mapping $\Gamma$ satisfies:
\proclaim Proposition 3.
It holds that
\begin{eqnarray}
(id \otimes \Gamma) \Delta (\Gamma (a)) & = & (\Gamma \otimes \Gamma
) \Delta (a)\,,
\cr
\Gamma (ab) & = & \Gamma (a) b ,\,\,\,\, a \in F_q(G)\,\,,
b \in {\cal C}\,.\label{im}
\end{eqnarray}

{\it  Proof}. The proof of the second equation is rather
straightforward using the decomposition of the universal element
${\rho}$. The proof of the first equation goes as follows. The
starting
point is
\begin{equation}
(id \otimes id \otimes \Gamma)(id \otimes \Delta){\cal Z} = (id
\otimes id \otimes \Gamma)({\cal Z}_1 \otimes \langle {\cal Z}_2 ,
{\rho}_1 {\tilde {\rho}}_1 \rangle {\rho}_2 \otimes {\tilde
{\rho}}_2)\,,\label{wich}
\end{equation}
where we used
\begin{eqnarray}
{\cal Z} &=& {\cal Z}_1 \otimes {\cal Z}_2\,,\cr
\rho &=& {\rho}_1 \otimes {\rho}_2\,,\cr
{\cal Z} &=& {\cal Z}_1 \otimes \langle {\cal Z}_2 , {\rho}_1 \rangle
{\rho}_2\,.
\end{eqnarray}
Applying the projection $\Gamma$ and using the decomposition of
$\rho$ yields
\begin{equation}
(id \otimes id \otimes \Gamma)(id \otimes \Delta){\cal Z} = {\cal
Z}_{12}{\cal Z}_{13}\,.
\end{equation}
Inserting $a$ in the first tensor factor gives the
result.\vspace{.5cm}\\
For later purposes we also introduce the dual mapping
\begin{equation}
\tilde{\Gamma} : \Uh({\gt g}) \rightarrow \Uh({\gt g}):
 x \mapsto (id \otimes \langle x,\cdot\rangle){\cal Z}\,.
\end{equation}
\section{Differential operators on ${\cal C}$}
The aim of the following paragraph is to introduce the partial
derivatives
${\partial \over\partial{Z^i_j}}$ with respect to the coordinates
$Z^i_j$  with
the help of the projection to ${\cal C}$ of the right action of
$\Uh$(\gtg) on
$F_q(G)$, and to express the action of the left-invariant vector
fields on
${\cal C}$ in terms of these.\vspace{.5cm}\\
{\bf {Definition 1.}}
\begin{equation}
 a \lhd X := \Gamma (a \ast X ) =\Gamma ( a_{(2)}) \langle X,
a_{(1)}\rangle,\quad a\in F_q(G), X\in \Uh(\gtg)\,.
\end{equation}\vspace{.5cm}
Using the explicit form of the mapping $\Gamma$, it is easy to see
that we can write:
\begin{equation}
a \lhd X = (\langle a, X \cdot\rangle \otimes id){\cal
Z}\,.\label{ract}
\end{equation}
The ``action'' $\lhd $ has the following properties:
\proclaim Proposition 4.
\begin{eqnarray}
a \lhd XY & = & \epsilon(X) a\lhd Y \, \quad \mbox{if}\,\, X = (id
\otimes
\langle X',\cdot\rangle )\Lambda\, , \, a\in {\cal C}\,,\cr
a \lhd XY &=& (a \lhd X) \lhd Y \,\quad \mbox{if}\,\, Y \in
Im{\tilde{\Gamma}}
\,.\label{action}
\end{eqnarray}

The above-defined action (it is really an action of
$Im{\tilde{\Gamma}}$ on
${\cal C}$ according to (\ref{action})) now serves to introduce a
complete set
of
partial differential operators on the space ${\cal C}$. In order to
do so we start from the following observation:
\begin{equation}
Z^a_b \lhd X = {\tau}^a_c (X) Z^c_b\,, \quad \mbox{where}\quad X\in
Im
{\tilde{\Gamma}}\,.
\label{dar}\end{equation}
To introduce an apropriate set of differential operators on ${\cal
C}$ we choose the following elements of $Im \tilde{\Gamma}$:
\begin{equation}
\beta := (q-q^{-1})^{-1} (\tau \otimes id)(R^u Q^{-1u}-1)\,.
\label{beta}
\end{equation}
{}From now on, we shall restrict ourselves to the case $G=SL(N)$ and
$P_0=B_-$,
where $B_-$ is the Borel subgroup of lower triangular matrices. In
this case
the matrix $Z$ is an upper triangular matrix with units on the
diagonal. The
$R$-matrices $R$ and $Q
$ are then of the form
\begin{eqnarray*}
q^{1/N}R^{jk}_{st} & = &
\delta^j_s\delta^k_t+(q-q^{\sgn(k-j)})\,\delta^j_t
\delta^k_s\, ,   \cr\cr
q^{1/N}Q^{jk}_{st} & = & q^{\delta^{jk}}\delta^j_s\delta^k_t\, ,
\cr\cr
\end{eqnarray*}
and the relation (\ref{comm}) can be rewritten, for the individual
matrix
entries, as
$$
q^{\delta^k_s}Z^j_sZ^k_t-q^{\delta^j_t}Z^k_tZ^j_s=
(q^{\sgn(k-j)}-q^{\sgn(s-t)})q^{\delta^j_s}Z^k_sZ^j_t\, .
\eqno(7.17) $$

Using (\ref{dar}) in the definition of $\beta$ yields in this case
the
following result for the right action of the
functionals (\ref{beta}) on the matrix of the holomorphic coordinates
on the homogeneous space $(P_0\backslash G)_q$,
\begin{equation}
Z^a_b \lhd {\beta}^i_j = {\delta}^a_j Z^i_b\quad\mbox{for}\quad i >
j\,.\label{der}
\end{equation}
This identity implies the following ansatz for the $\beta$'s in terms
of derivatives in the variables $Z$
\begin{equation}
a \lhd \beta^i_j = Z^i_r {\partial a \over\partial{Z^j_r}}.\label
{ansatz}
\end{equation}
Therefore we define the partial derivatives on the space ${\cal C}$
through the action of the functionals $\beta$. In order to obtain a
complete
description of the partial derivatives ${\partial
\over\partial{Z^i_j}}$ we
have to specify the deformed Leibnitz rule.
This is done by starting from the comultiplication of the $\beta$'s.
Using
\begin{eqnarray}
(\Delta \otimes id) Q^u &=&  {Q^u}_{13} {Q^u}_{23}\,,\cr
Q^u &= &{Q^u}_{21}\,,\label{q}
\end{eqnarray}
one obtains the following comultiplication of the $\beta$'s:
\begin{equation}
\Delta ({\beta}^i_j) = S({L^-}^k_m ){Q^{-1}}^r_j \otimes S({L^-}^i_k)
{Q^{-1}}^m_r - {\delta}^i_j 1 \otimes 1\,.
\end{equation}
In order to derive the deformed
Leibnitz rule for the derivatives $\beta$, we make use of the
following observation:
\begin{equation}
\Gamma (a) \lhd X = a \lhd \tilde{\Gamma} (X)\,.
\end{equation}
So in the case where $X$ is already an element of $Im\tilde{\Gamma}$
the
right action on any element of $F_q(G)$ is identical to the
right action on its projection.
Therefore using the first equation in (\ref{im}) and the definition
of
the matrix $Z$ one obtains for $f \in {\cal C}$
\begin{equation}
(Z^a_b f) \lhd {\beta}^i_j = (T^a_b f)\lhd {\beta}^i_j\,.
\end{equation}
Starting from this equation we finally end up with the following
Leibnitz rule on $\cal C$:
$$(Z^a_b f) \lhd {\beta}^i_j = (Z^a_b \lhd {\beta}^i_j) f +
q^{(-\delta^{ia}+\delta^i_b-\delta_{jb})}Z^a_b (f\lhd {\beta}^i_j)$$
\begin{equation}
+ \delta^a_jq^{\delta^i_b}(q-q^{sgn(a-c)})Z^c_b(f\lhd {\beta}^i_c)\,.
\end{equation}
So (\ref{der}) together with the Leibnitz rule completely define the
derivatives on $\cal C$. Therefore it is now possible to express the
left action on ${\cal C}$ of the vector fields on the quantum group
through
functions
of the holomorphic coordinates $Z$ and the differential operators
$\beta$
(or ${\partial  \over\partial{Z^i_j}}$).\vspace{.5cm}\\
The vector fields are defined in the standard way \cite{Jurco},
\cite{S-W},
\cite{Schupp}
\begin{equation}
\kappa:= (q-q^{-1})^{-1} (\tau \otimes id) (1 - R^u_{21}
R^u)\,.\label{vector}
\end{equation}

\proclaim Proposition 5.
On the space ${\cal C}$ we obtain the following representation of the
vector fields
\begin{equation}
{\kappa}^i_j \ast a = - Z^{-1i}_k Z^l_j q^{-{\delta}^l_j}
q^{{\delta}^k_j}
Z^k_r  {\partial a \over\partial{Z^l_r}}\,,\quad\,k>l\,.
\end{equation}

{\it Proof}. Using (\ref{comm}) it is straightforward to derive the
following commutation relation
\begin{equation}
{\cal Z}_{23} R^u_{21} R^u_{12} = Q^{-1u}_{12} {\cal Z}^{-1}_{13}
Q^u_{12}
R^u_{21} R^u_{12} Q^{-1u}_{12} {\cal Z}_{13} Q^u_{12} {\cal
Z}_{23}\,.\label{lr}
\end{equation}
Now using

\begin{equation}
(id \otimes X* ) \rho = \rho (X \otimes id )\,,
\end{equation}
and the commutation relation (\ref{lr}) it is possible to derive the
following identity on $a \in {\cal C}$ by applying $( {\tau}^i_j
\otimes a \otimes id)$ on both sides of (\ref{lr})
\begin{equation}
{\kappa}^i_j \ast a = - Z^{-1i}_k Z^l_m \,a \lhd ( {\beta}^k_l Q^m_j
)\,.\label{id}
\end{equation}
To bring this equation to the desired form, we have to commute $Q$
and $\beta$
and use (\ref{action}).
Starting from (\ref{q}) and the fact that $\Uh({\gt g})$ is
quasitriangular, one obtains the following commutation relation (on
$a \in {\cal C}$)

\begin{equation}
a \lhd [{\beta}^i_m Q^{mk}_{jr} Q^r_l - Q^k_r Q^{ir}_{ml}
{\beta}^m_j] = 0\,.\label{betq}
\end{equation}
Therefore inserting (\ref{betq}) in (\ref{id})  yields the expression
in Proposition 5 for the left action of the vector fields on ${\cal
C}$
in terms of holomorphic coordinates and derivatives.
\vskip 0.5cm
\noindent{\bf Remark:}
It is easy to see that, for $a\in {\cal C}$:
$${\kappa}^i_j*a=({\kappa}^i_j*Z^m_n){\partial a
\over\partial{Z^m_n}}.$$
So if we assume the differential calculus on $SL_q(N)$ with the
conventions of,
for instance \cite{J-St},
we have for the differential of $a\in{\cal C}$
$$da=dZ^m_n{\partial a \over\partial{Z^m_n}}.$$
This observation justifies our ansatz (\ref{ansatz}) for partial
derivatives.

\section{Representations}
Using the proof of Proposition 5, one can now easily generalize the
above formulas to the case of the action of ${\kappa}^i_j$ in an
arbitrary
irreducible finite-dimensional representation \Tl of  $\Uh$({\gt
sl}(N))
corresponding
to a maximal weight $\lambda=(m_1,m_2,...,m_{N-1})$ \cite{Ro},
\cite{Lu},
\cite{P-W}. First let us embed the representation space \Hl
into ${\cal C}$ by $| \Psi \rangle\ \mapsto \Psi_{\lambda}:=\langle
\lambda |{\cal Z} | \Psi \rangle$. So it is now clear that
instead of applying the second component of the triple tensor product
in
(\ref{lr}) to an element $a\in{\cal C}$ in going from (\ref{lr}) to
(\ref{id}) we have now to compute the matrix element $\langle
\lambda |.| \Psi \rangle$ of the second component in order to obtain
the action
of ${\kappa}^i_j$ on $\Psi_{\lambda}$. We use the notation
$m_i=\tilde
m_i-\tilde m_{i+1}$, $\sum \tilde m_i =0$.

The result is

\proclaim Proposition 6.
The action of the ${\kappa}^i_j$ in the representation \Tl
in terms of holomorphic coordinates and
derivatives is  given by
\begin{eqnarray}
({\kappa}^i_j * {\Psi}_{\lambda})(Z)& = &{(q - q^{-1})}^{-1}
q^{-{\tilde m}_i}q^{{\tilde m}_k}q^{{\tilde m}_j}( q^{-{\tilde m}_k}
- q^{{\tilde m}_k}) {Z^{-1}}^i_k Z^k_j {\Psi}_{\lambda}(Z) \nonumber
\\ & & - q^{{\tilde m}_k} q^{{\delta}^k_j} q^{-{\tilde m}_i}
q^{{\tilde m}_k} q^{-{\delta}^l_j} q^{{\tilde m}_j} {Z^{-1}}^i_k
Z^l_j ({\Psi}_{\lambda}(Z) \lhd {\beta}^k_l)\,.
\end{eqnarray}

{\it Proof}. Inserting ${\tau}^i_j$ in the first component of the
tensor product (\ref{lr})
we obtain the following expression
\begin{equation}
{\cal Z}_{12} (({\tau}^i_j \otimes id) R_{21} R_{12})_{1} =
({Q^{-1}}^i_k Q^l_m (({\tau}^i_j \otimes id) R_{21} R_{12})
{Q^{-1}}^n_o Q^p_j)_1
 ({Z^{-1}}^k_l Z^o_p)_2 {\cal Z}_{12}\,.
\end{equation}
The restriction to the specific state $\Psi$ in the representation
\Tl is done by
taking the matrix element between $\langle \lambda| $ and $| {\Psi}
\rangle$ in the first tensor factor (all matrix elements in the
following are in the first tensor factor), where $\langle \lambda| $
is defined by the following equations:
\begin{eqnarray}
Q^i_j &= &{\delta}^i_j q^{{\tilde H}_i}\,,\cr
{\tilde H}_i - {\tilde H}_{i+1} &= &H_i\,,\cr
\sum_{i} {\tilde H}_i& = &0\,,\cr
{\tilde H}_i |\lambda \rangle &=& {\tilde m}_i |\lambda \rangle\,,\cr
\langle \lambda |{ L^{+}}^i_j &=& \langle \lambda | {\delta}^i_j
q^{{\tilde m}_i}\,.\label{repr}
\end{eqnarray}
This yields

\begin{eqnarray}
{\langle \lambda |}{\cal Z}_{12} ({\kappa}^i_j)_{1} |{\Psi}\rangle &
= &q^{-{\tilde m}_i}
q^{{\tilde m}_k} Z^{-1i}_k Z^l_n [{(q-q^{-1})}^{-1}( q^{-{\tilde
m}_k} - q^{{\tilde m}_k} ) {\delta}^k_l \langle \lambda| Q^n_j {\cal
Z}_{12} | {\Psi} \rangle \nonumber \\
& & - q^{{\tilde m}_k} \langle \lambda | {\beta}^k_l Q^n_j {\cal
Z}_{12} | \Psi \rangle ]\,,
\end{eqnarray}
where the ${\kappa}$ are the vector fields defined in (\ref{vector}).

Using the evaluation of $Q$ on $\langle \lambda|$ as defined in
(\ref{repr}) and the commutation relations between $\beta$ and $Q$
defined in (\ref{betq}) one obtains
\begin{eqnarray}
{\langle \lambda |}{\cal Z}_{12} (X^i_j)_{1} |{\Psi}\rangle & = &
-q^{{\delta}^k_j}
q^{{\delta}^l_j} q^{-{\tilde m}_i} q^{{\tilde m}_k} Z^{-1i}_k Z^l_j
\langle \lambda| {\beta}^k_l {\cal Z} | \Psi \rangle \nonumber\\ && +
{(q-q^{-1})}^{-1} q^{-{\tilde m}_i}q^{{\tilde m}_k}q^{{\tilde m}_j}
(q^{-{\tilde m}_k} - q^{{\tilde m}_k}) Z^{-1i}_k Z^k_j \langle
\lambda |{\cal Z} | \Psi \rangle\,.
\end{eqnarray}

Identifying
\begin{equation}
\langle X, {\Psi}_{\lambda} \rangle = \langle \lambda| X | {\Psi}
\rangle\,,\quad X\in Im\tilde\Gamma\,,
\end{equation}
and using the definition of the left and right action $\lhd$ (in
the form given in (\ref{ract})) we obtain the desired
result.\vspace{1cm}\\
If we now assume that the generators $\kappa$ are acting on the whole
${\cal
C}$, we can interpret our results as a natural Fock space
representation of the universal enveloping algebra of $\Uh$(\gtg)
($\Uh$(\gtk))
as the action is expressed in terms of holomorphic
coordinates $Z^i_j$ and derivatives ${\partial
\over\partial{Z^i_j}}$. These representations are nothing but the
contragradient Verma modules.\vspace{.5cm}\\

{\it Acknowledgement.} The authors would like to thank Bruno Zumino
for many
valuable discussions.

\end{document}